\documentclass[aps,prl,twocolumn,floatfix,showpacs]{revtex4}
\usepackage{graphicx}
\usepackage{verbatim}
\begin{document}
\title{ High pressures make Hg a transition metal in a thermodynamically stable solid}
\author{Xiaoli Wang$^{1,2}$}
\author{Haiqing Lin$^{1,3}$}
\author{Yanming Ma$^{4,1}$} 
\author{Mao-sheng Miao$^{5,1}$}
\email{miaoms@mrl.ucsb.edu}
\affiliation{$^{1}$Beijing Computational Science Research Center, Beijing 10084, P. R. China}
\affiliation{$^{2}$Institute of Condensed Matter Physics, Linyi University, Linyi 276005, P. R. China}
\affiliation{$^{3}$Department of Physics and Institute of Theoretical Physics, Chinese University of Hong Kong, Hong Kong SAR, P.R. China}
\affiliation{$^{4}$State Key Lab of Superhard Materials, Jilin University, Changchun 130012, P.R. China}
\affiliation{$^{5}$Materials Department and Materials Research Laboratory, University of California, Santa Barbara, CA 93106-5050}

\begin{abstract}
The appropriateness of including Hg among the transition metals
has been debated for a long time. 
Although the synthesis of HgF$_{4}$ molecules in gas phase 
was reported before, the molecules show strong instabilities and dissociate. 
Therefore, the transition metal propensity of Hg remains an open question. 
Here, we propose that high pressure provides a controllable method for preparing
unusual oxidation states of matter. Using an advanced structure search method based on first-principles electronic structure
calculations, we predict that under high pressures, Hg can transfer the electrons in
its outmost $d$ shell to F atoms, thereby acting as a transition metal. Oxidation of 
Hg to the +4 state yielded thermodynamically stable molecular
crystals consisting of HgF$_{4}$ planar molecules, a typical geometry 
for $d^{8}$ metal centers.
\end{abstract}


\maketitle

A goal of physics and chemistry is to prepare unusual states of
matter beyond the naturally occurring forms. One of the most challenging and attractive
topics in this field is the preparation of high oxidation states of
Hg.\cite{Wang2007,Jensen2008} Group IIB elements, including Zn, Cd, and Hg are
usually defined as post-transition metals\cite{IUPAC-gold} because they are commonly
oxidized only to the +2
state.\cite{Greenwood1984,Cotton1999} Their $d$ shells are completely filled
and do not participate in the formation of chemical bonds; however, 
there is a high expectation that Hg should be stable at higher oxidation states due to  the
large relativistic effects that increase the energies of its $5d$ levels.

Earlier attempts focused on HgF$_{4}$ molecules in gas phases, including  
several quantum chemistry calculations on the molecule.\cite{Jorgensen1979,Jorgensen1986,Riedel2004,Liu1999,Pyykko2002,Deming1976,Hrobarik2008,Riedel2005}
It was predicted that HgF$_{4}$ molecules can resist dissociation to HgF$_{2}$ and F$_{2}$ molecules.
\cite{Kaupp1993,Kaupp1994,Riedel2004} However, the thermodynamically
stable form of HgF$_{2}$ at low temperatures is the solid phase rather than the gas
phase.\cite{Jensen2008} HgF$_{4}$ molecules in the gas phase or in molecular crystals can, therefore, become unstable  and
dissociate to form HgF$_{2}$ and F$_{2}$, as predicted by our calculations (see
below). For this reason, the synthesis of HgF$_{4}$ molecules has shown to be
extremely difficult. Recently, after numerous experimental attempts, 
the trace of HgF$_{4}$ molecules was observed using matrix-isolation infrared
(IR) spectroscopy.\cite{Wang2007} The HgF$_{4}$ molecules were unstable and dissociated to yield HgF$_{2}$ and
F$_{2}$ within very short periods of time. To demonstrate that Hg is truly a transition 
metal, one must \emph{thermodynamically} stabilize a Hg center in the +4 oxidation state.

\begin{figure}
\includegraphics[width=9cm]{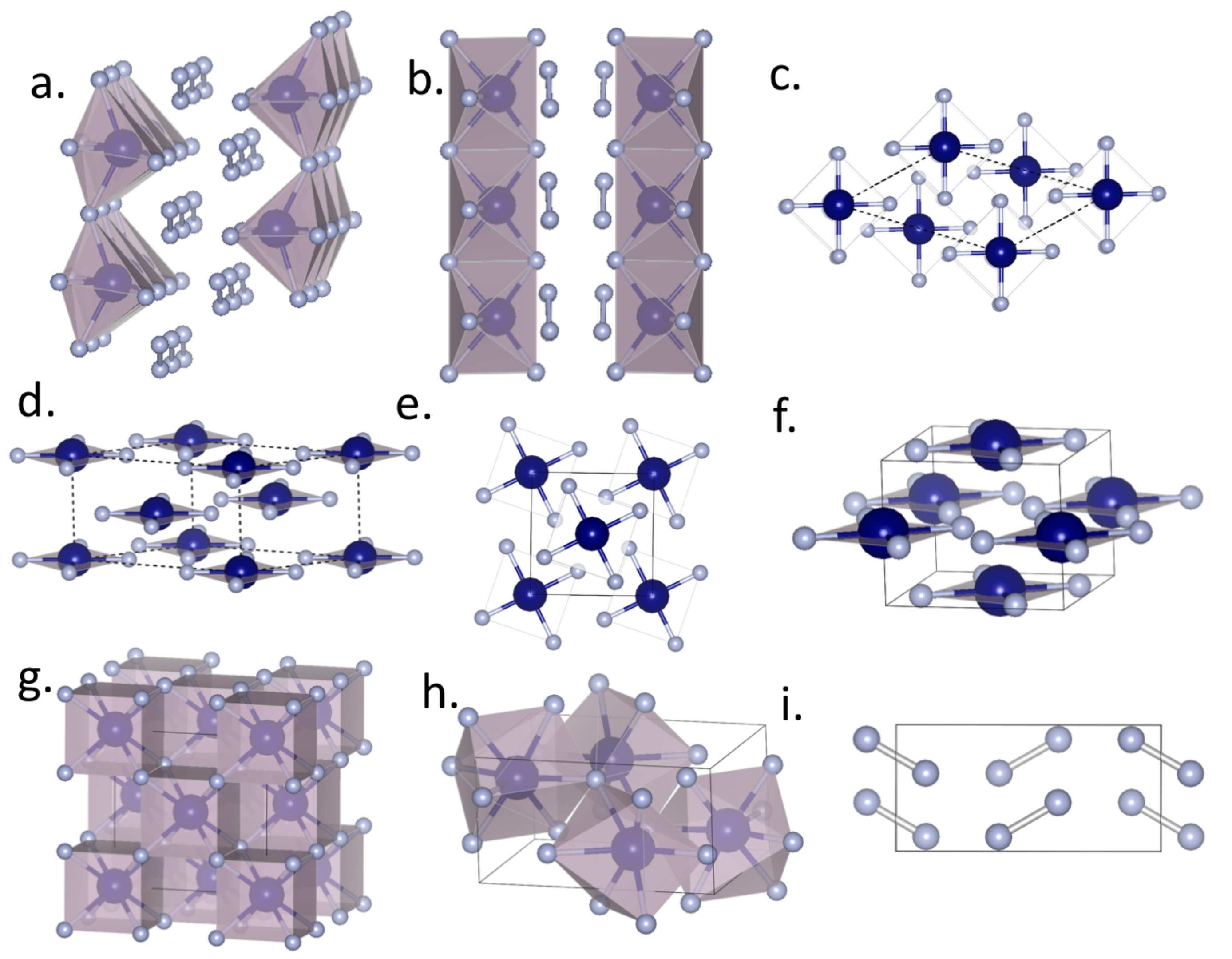}
\caption{\footnotesize (Color figures available online.) (a) and (b): Two views of the
$Pca2_{1}$ structure, the optimized lattice parameters at 15 GPa are a = 14.183 , b = 3.478  and c = 4.153  with Hg occupying 4a (0.082, 0.277, 0.486) and F occupying 4a (0.309, 0.530, 0.321), 4a (0.306, 0.173, 0.523), 4a (0.663, 0.211, 0.337) and 4a (0.521, 0.771, 0.982) positions; (c) side view and (d)  top view of the $I4/m$
structure, the optimized lattice parameters at 150 GPa are a = 4.366  and c = 3.498  with Hg occupying 2b (0, 0, 0.500) and F occupying 8h (0.304, 0.888, 0) positions; (e) top view and (f) side view of the $I4/mmm$ structure; (g)
the CaF$_{2}$ structure; (h) the HgCl$_{2}$ structure; (i) the F$_{2}$
molecular crystal structure. The large dark-blue balls and the small
light-grey balls indicate Hg and F atoms, respectively. Parameters of more related structures can be found in the online supplementary materials. \label{structures} }
\end{figure}

In this letter, we propose a totally new approach that utilizes high-pressure
techniques to stabilize Hg$^{4+}$ in the solid phase. Over the past
several decades, high-pressure methods based on diamond anvil cells (DACs) have 
yielded novel structures of elements and compounds
under extremely high pressures (beyond megabars). This field has been
particularly advanced by the recent development of first-principles
structure-searching methods that can predict high-pressure structures to
extremely high accuracy without experimental or intuited input.  For
example, the recently implemented particle swarm optimization (PSO) technique
\cite{Wang2010,Ma2010} successfully predicted the high-pressure
structures of various systems, including lithium\cite{Lv2011}, water
ice\cite{Wang2011}, and Bi$_{2}$Te$_{3}.$\cite{Zhu2011} In most high-
pressure studies, the oxidation states of the compounds remain the same at high
pressures. We show here that the high-pressure techniques have the potential to 
provide a controllable method for achieving unusual oxidation states.

We conducted a structure search over
the full structure configuration space, based on {\it ab initio} total-energy
calculations and a PSO technique, implemented
in the CALYPSO (crystal structure analysis by particle swarm optimization)
code.\cite{Wang2010,Ma2010} The underlying {\it ab initio} structural
relaxations and electronic band structure calculations were performed using the
framework of density functional theory (DFT), implemented in the VASP
code.\cite{Kresse1996} The generalized gradient approximation (GGA) in the
framework of Perdew--Burke--Ernzerhof (PBE)\cite{Perdew1996} was used to treat the
exchange-correlation functional, and the projector augmented wave
(PAW)\cite{PAW} pseudopotentials were used to describe the ionic potentials. The
cutoff energy for the expansion of the wave function into plane waves was set
to 700 eV, and the Monkhorst--Pack $k$ meshes \cite{Monkhorst1976} were chosen to
ensure that all enthalpy calculations converged to better than 1
meV/atom. The structure searches were performed with a unit cell containing
20 atoms (4 Hg and 16 F) and were performed over the pressure range 0--300
GPa. The phonon spectra were calculated using the Phonopy program.\cite{phonon}

\begin{figure}[tbp]
\includegraphics[width=8.5cm]{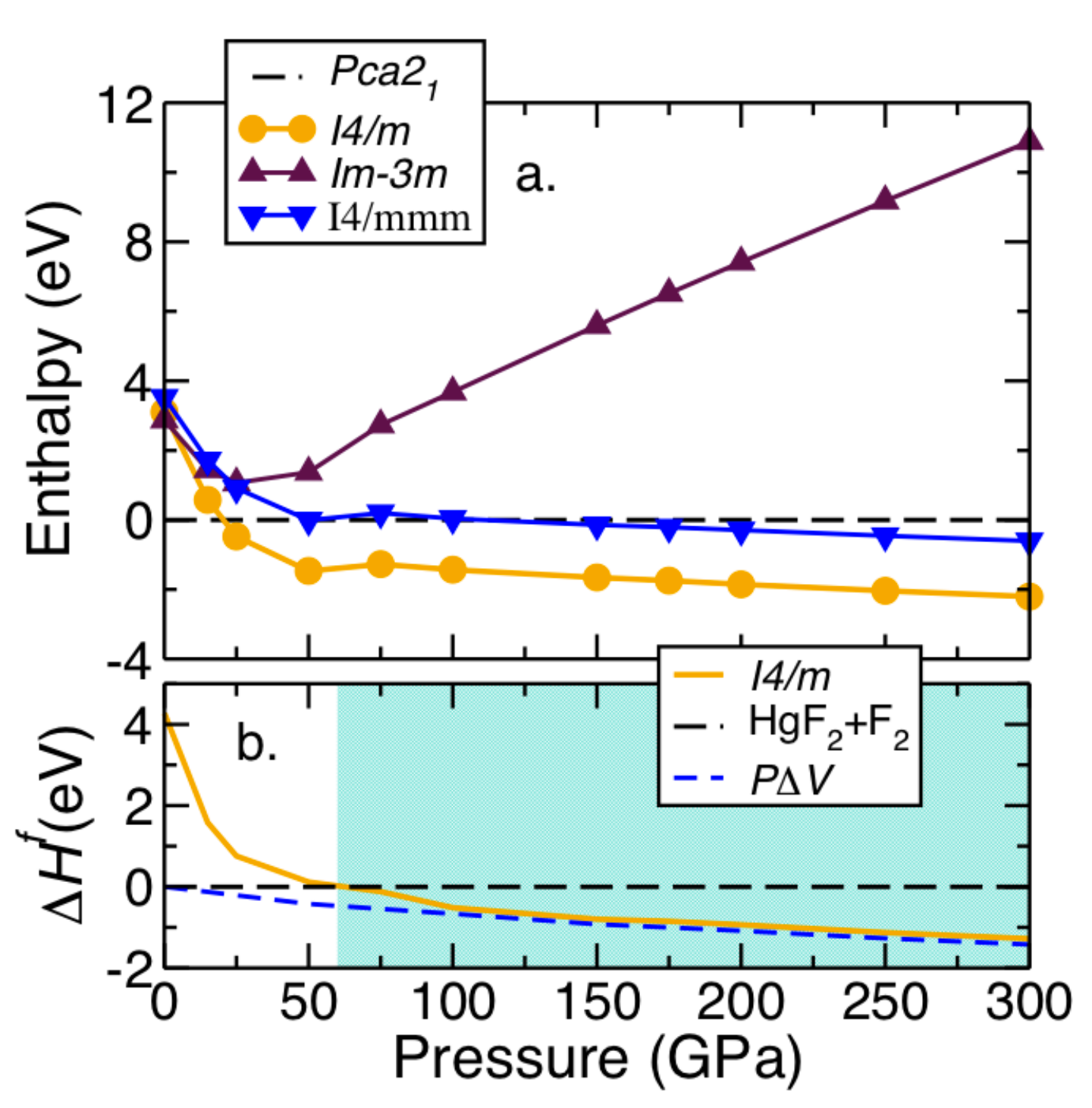}
\caption{\footnotesize (Color figures available online.) (a) Comparison of the enthalpies as
a function of the pressure for the Hg$_{1}$-F$_{4}$ structures selected from the PSO
search. The results are reported with respect to the enthalpy of the $Pca2_{1}$ structure,
which is the most stable structure under ambient conditions. (b) Energy of the
HgF$_{2}$(s)+F$_{2}$(s)$\rightarrow$ HgF$_{4}$ (s) reaction. The enthalpy of
HgF$_{4}$ was obtained from the $I4/m$ structure. The blue dashed line shows the
enthalpy gain due to the volume reduction, $P\Delta V$. The area shaded in light
blue indicates the range of pressures under which HgF$_{4}$ is thermodynamically
stable at 0 K.  \label{enthalpy} } \end{figure}

As shown in Fig. \ref{enthalpy}(a), two thermodynamically
stable structures were identified over the pressure range of 0--300 GPa. The
most stable structure under ambient conditions had a symmetry of $Pca2_{1}$
and consisted of chain-like HgF$_{2}$ and F$_{2}$ molecules [Fig.\ref{structures}(a) and (b)],
indicating that at low pressures, the Hg$_{1}$-F$_{4}$ system tended to decompose
into HgF$_{2}$ and F$_{2}$. The shortest F-F distances were 1.485 \AA~ at 0 GPa,
similar to the F-F bond length in the gas phase.\cite {Chase1985} At a
pressure of 20 GPa, the system transformed into a structure with $I4/m$
symmetry [Fig.\ref{structures}(c) and (d)]. This structure consisted of HgF$_{4}$ molecules and 
each Hg had four nearest neighboring F atoms. At 50 GPa, the Hg-F distance was 1.949 \AA. 
This distance was significantly shorter than the average Hg-F distance of 2.215 \AA~ at the same pressure. 
On the other hand, the shortest F-F distance was 2.112 \AA~ which was significantly longer than the F-F molecular
bond length of 1.439 \AA, indicating that the molecular features of F$_{2}$ were destroyed upon
formation of the HgF$_{4}$ molecules. The structural parameters for both structures and 
the dependence of bond lengths in $I4/m$ 
structure on increasing pressure can be found in the online supplemental material. 

Another structure displayed a symmetry of $I4/mmm$ [Fig.
\ref{structures}(e) and \ref{structures}(f)]. Although this structure appeared similar to the
$I4/m$ structure as viewed from the top, each Hg atom was bonded to eight F atoms to form a
chain. The enthalpy of this structure was lower than that of the
$Pca2_{1}$ structure at ~100 GPa; however, over the pressure range 0--300
GPa, the enthalpy exceeded that of the $I4/m$ structure, a HgF$_{4}$
molecular crystal. 

As shown experimentally\cite{Poole1979,Hostettler2005} and in our calculations, HgF$_{2}$ was stable in extended
solid structures other than the HgF$_{2}$ chains present in the
$Pca2_{1}$ structure of the Hg$_{1}$-F$_{4}$ system. To determine whether the
separated HgF$_{2}$ and F$_{2}$ phases could react to form a HgF$_{4}$ phase, we
calculated the energy of the HgF$_{2}$(s)+F$_{2}$(s)$\rightarrow$ HgF$_{4}$ (s)
reaction. The structures of HgF$_{2}$ and F$_{2}$ under pressure were
searched using the PSO method. HgF$_{2}$ was found to be stable
in the CaF$_{2}$ structure at 0 to 100 GPa,\cite{Poole1979,Hostettler2005} 
in the low-pressure structure of HgCl$_{2}$ ($Pnma$)\cite{Subramanian1980} from 100 to 300 GPa. 
We also found that F$_{2}$ maintained its molecular features and was
stable in its  low-pressure structure $C12/c1$\cite{Pauling1970} up to 300 GPa. 

As shown in Fig. \ref{enthalpy}(b), our results clearly showed that at ambient
pressures, the separated HgF$_{2}$ and F$_{2}$ phases were stable against the formation
of HgF$_{4}$, {\it i.e.} the reaction energy was positive. The reaction
energy decreased rapidly as the pressure increased and becomes negative at 
60 GPa, indicating that HgF$_{4}$ became thermodynamically stable. The energy
gain upon formation of the HgF$_{4}$ molecules was quite large and increased
significantly with the increasing pressure. At a pressure of 150 GPa, the energy
gain was 0.8 eV or 77 kJ/mole, providing a large driving force for oxidation of Hg to the +4
state.

\begin{figure}
\includegraphics[width=8.5cm]{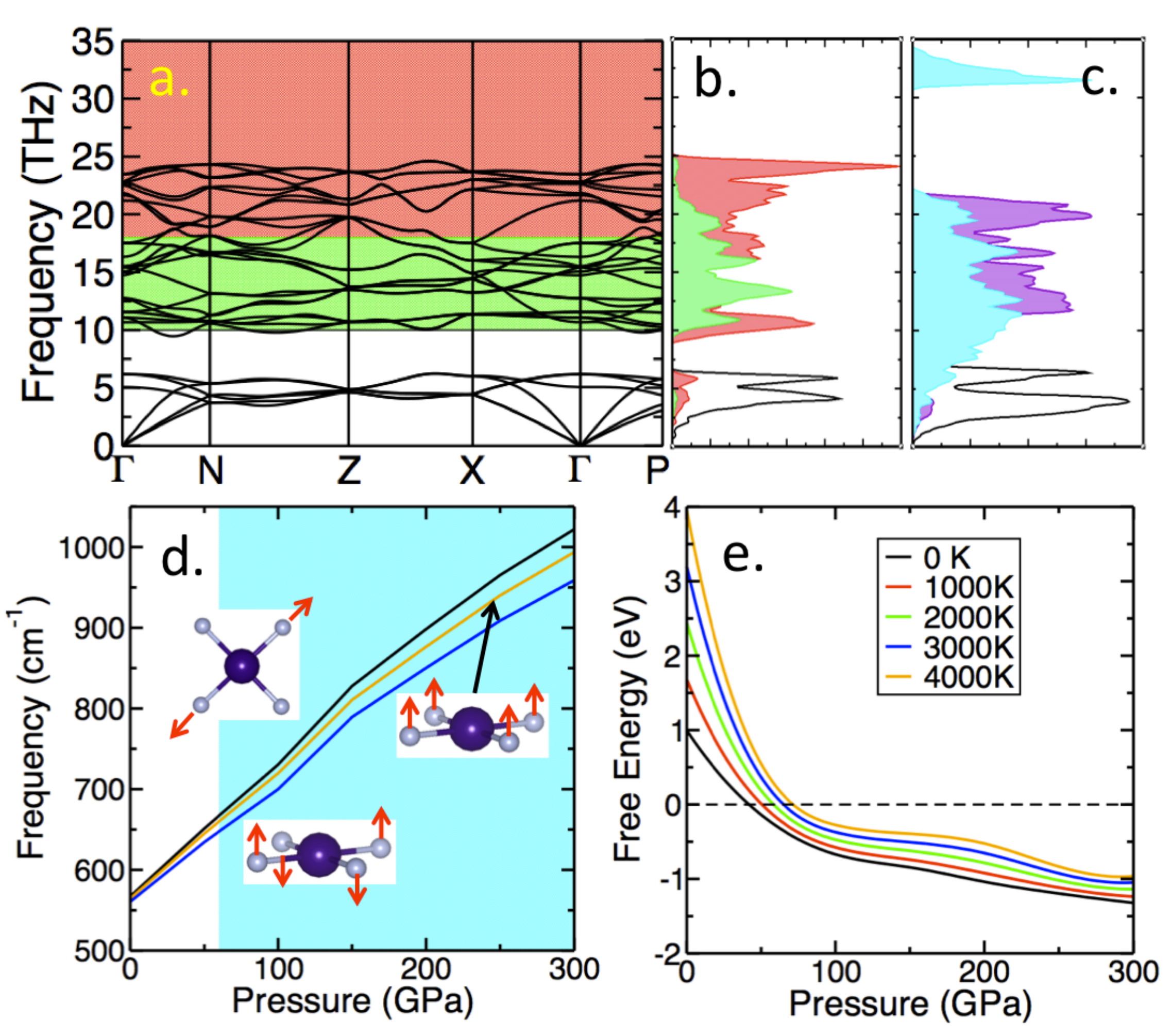}
\caption{\footnotesize (Color figures available online.) (a) Phonon spectrum of the HgF$_{4}$
molecular crystal in the $I4/m$ structure under a pressure of 150 GPa. The energy regions that
corresponded to the F in-plane, the F out-of-plane, and the Hg motions are
shaded in red, green, and white. (b) The phonon projected density of states (PDOS) of HgF$_{4}$ in the
$I4/m$ structure at a pressure of 150 GPa. The black line and the red and green shaded
areas show the DOS projected onto the Hg, the F in-plane, and the F out-of-plane
motions. (c) The summed phonon PDOS of HgF$_{2}$ in the $Pnma$ structure and
F$_{2}$ in the $C12/c1$ structure at 150 GPa. The black line shows the motions of Hg
in HgF$_{2}$. The blue and purple shaded areas show the motions of F in the
HgF$_{2}$ and F$_{2}$ solids, respectively. (d) The three selected $I4/m$
phonon frequencies at $\Gamma$ as a function of the pressure. The black, 
orange, and blue lines denote the Hg-F stretching and bending modes, as shown
in the insets. (e) The calculated reaction free-energies as a function of
the pressure at 0, 1000, 2000, 3000, and 4000 Kelvin.  \label{phonon} }
\end{figure}

The phonon spectrum of HgF$_{4}$ in the $I4/m$ structure, of HgF$_{2}$ in the $Pnma$
structure, and of F$_{2}$ in the $C12/c1$ structure were calculated over the
pressure range of 0--300 GPa. The results show that the structures of all
three solids were dynamically stable up to 300 GPa, {\it i.e.} none of the
phonon modes featured imaginary frequencies. 
As shown in Fig. \ref{phonon}(a) and (b), the low-frequency modes of the HgF$_{4}$
molecular crystal at 150 GPa (from 0 to 10 THz) were dominated by the motions of the
Hg atoms, whereas the modes in the frequency range 10--18 THz and 18--%
25 THz were dominated by the F out-of-plane (z) and in-plane (x and y)
motions. The Hg-F stretching modes in HgF$_{4}$ reached their peaks at 24 THz (800
cm$^{-1}$). On the other hand, the Hg-F stretching modes in HgF$_{2}$ reached their peak
around 20 THz (650 cm$^{-1}$), as shown in Fig. \ref{phonon}(c). The F-F
stretching modes were much higher in energy, and a frequency gap was observed from
21 to 31 THz (700--1000 cm$^{-1}$) in a HgF$_{2}$ and F$_{2}$ mixed sample. The
vibrational peak observed in this gap will indicate the formation of HgF$_{4}$
molecular crystals. The HgF$_{4}$ vibrational modes in this frequency gap included
Hg-F stretching and bending modes and could be either IR or Raman active.
As shown in Fig. \ref{phonon}(d), the frequencies of these peaks tended to split under higher
pressures, indicating a broadening of the HgF$_{4}$ signature peaks with increasing
pressure. 

Another important issue associated with the synthesis of HgF$_{4}$ in a high-pressure DAC is
the persistence of the stable structure at high temperatures. Although HgF$_{4}$ is
thermodynamically stable at 0 K, it does not form easily at very low temperatures
in a DAC because of the large kinetic barrier to the solid-state
reaction. To overcome this barrier, thermo-heating or laser-heating
techniques may be used to increase the temperature in the DAC up to a few thousand degrees
Kelvin. We calculated the reaction energies at finite temperatures using the
calculated phonon spectra of the HgF$_{4}$, HgF$_{2}$, and F$_{2}$ solids. As shown
in Fig. \ref{phonon}(e), elevated temperatures tended to destabilize the HgF$_{4}$
solid; however, even at $T=4000$ K, HgF$_{4}$ could still reach thermodynamic
stability at pressures exceeding 75 GPa. 

\begin{figure}[tbh] \hspace{1cm}
\includegraphics[width=8.5cm]{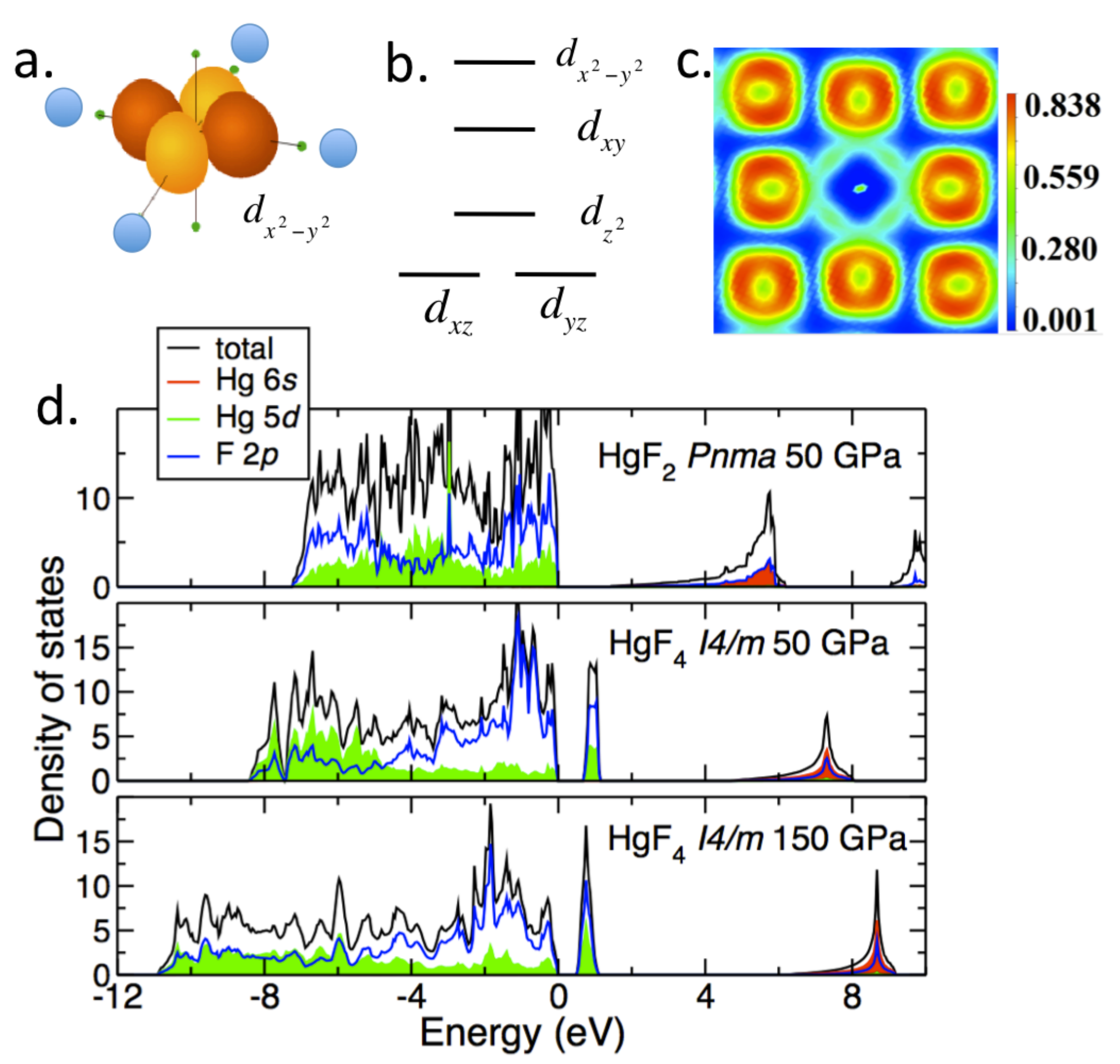}
\caption{\footnotesize (Color figures available online.) (a) The crystal field splitting diagram of a square planar complex;
the light-blue balls indicate the F anions and the orange and red spheres
indicate the $d_{x^{2}-y^{2}}$ orbit; (b) the $d$ orbitals were split under the square
planar crystal field; (c) the calculated electron localization function (ELF)
for HgF$_{4}$ in the $I4/m$ structure at 150 GPa; (d) the calculated PDOS for
HgF$_{2}$ in the HgCl$_{2}$ structure at 50 GPa and for the HgF$_{4}$ molecular crystal
in the $I4/m$ structure at 50 and 150 GPa. In the PDOS plots, the black solid lines indicate
the total DOS, the blue solid lines indicate the F $2p$ states, the red and 
green shaded areas indicate the Hg $6s$ and $5d$ states.  \label{pdos} }
\end{figure}

The planar geometry of the HgF$_{4}$ molecules is typical of $d^{8}$
configurations, indicating the involvement of $5d$ electrons in the formation of Hg-F
bonds. As shown in Fig. \ref{pdos}(a) and (b), the $d$ orbitals split into four
groups in a square planar crystal field due to differences in the
interactions with the four surrounding F anions. Among these orbitals, $d_{x^{2}-y^{2}}$ was the
highest in energy. The transition metal cations could be energetically stabilized
in this configuration in the event that 8 electrons filled the 4 lower $d$
states and left the $d_{x^{2}-y^{2}}$ states empty. The bonding features of HgF$_{4}$ were also examined  by
calculating the electron localization function (ELF)\cite{ELF} using the VASP code.
The results [Fig. \ref{pdos}(c)] clearly showed the equivalent covalent bonds
of Hg and its four neighboring F atoms.

More direct evidence for the involvement of $d$ electrons in the formation of chemical
bonds with F was obtained from the projected density of states (PDOS). We
calculated and compared the total DOS and PDOS for HgF$_{4}$ in the $I4/m$
structure and HgF$_{2}$ in the HgCl$_{2}$ structure at 50 GPa. HgF$_{2}$ revealed
a gap of 1.5 eV. The conduction bands consisted mainly of the F $p$ and Hg
$6s$ states, whereas the valences bands consisted of both F $p$ and Hg $5d$ states
[Fig. \ref{pdos}(d)]. The PDOS showed that all $5d$ states were filled, and Hg was
apparently in the +2 state. On the other hand, HgF$_{4}$ displayed a much smaller gap of
only 0.71 eV. More importantly, the $5d$ states were distributed over both the
valence and the conduction bands, indicating that the $5d$ electrons were
depleted, in agreement  with the planar geometry of the HgF$_{4}$ molecules and
the $d^{8}$ configuration analysis.  \\

Several effects of the high pressure assist in stabilizing the HgF$_{4}$ molecules. One mechanism
is based on the crystal field splitting. Because high pressures reduce the inter-atomic
distances, including those between Hg and its neighboring F atoms, the high pressures
enhance the crystal field splitting. As a result, the energy of the
$d_{x^{2}-y^{2}}$ state increases with increasing pressure. The PDOS of
HgF$_{4}$ in the $I4/m$ structure at 150 GPa is shown in Fig. \ref{pdos}(d).
A comparison with the PDOS at 50 GPa shows that although the gap changed only
slightly with increasing pressure, the center of the F $2p$ states was
significantly lower relative to the $d_{x^{2}-y^{2}}$ state, indicating a
larger energy gain with electron transfer from the $5d$ to the $2p$ orbital at higher pressures.
Another important factor is the reduction of the volume during the formation of solid
HgF$_{4}$. This factor further increases the enthalpy gain. As shown in Fig.
\ref{enthalpy}(b), the value of $P\Delta V$ increased with increasing pressure.
Although its contribution to the enthalpy gain was small at low pressures, this contribution
became dominant at pressures exceeding 50 GPa. The reduction in the volume
arose from the shortening of Hg-F bonds as Hg changed from the +2 to the +4 oxidation
states due to the involvement of the $5d$ electrons in the formation of stronger
Hg-F bonds with a larger covalent component. When the enthalpy gain was sufficiently large,
the reaction HgF$_{2}$(s)+F$_{2}$(s) $\rightarrow$ HgF$_{4}$ (s)
became exothermic and HgF$_{4}$ became thermodynamically stable. As calculated,
this occurred at 60 GPa.

In summary, we demonstrated that under an external pressure applied using
the DAC technique, Hg could be thermodynamically stabilized in a high oxidation
state to form solid structures of HgF$_{4}$ molecules. The atomic
structure and the electronic structure of the high-pressure phases suggested that the
$5d$ orbitals of Hg were involved in chemical bonding. Thus, our calculations suggested
that Hg behaves as a true transition metal under high pressures. High-pressure techniques have been used to alter the chemical inertness of
noble gases and to sustain them in unusual chemical
states.\cite{Japhcoat1998,Somayazulu2010,Caldwell1997} Our study
also suggests that high-pressure techniques may be used as a controllable method
to achieve unusual oxidation states in matter.  \\

{\bf Acknowledgement} MSM thanks Prof. Ram Seshadri for inspiring discussions on the high oxidation
states of transition metals. MSM also thanks the ConvEne-IGERT Program (NSF-DGE
0801627). XW was partially supported by the National Natural Science Foundation
of China (Grant N 11147007). YM was supported by the National Natural Science
Foundation of China (Nos. 11025418 and 91022029)

\end{document}